\documentclass[sigconf]{acmart}
\usepackage{multirow}
\usepackage{times}

\usepackage{soul}
\usepackage{url}
\AtBeginDocument{%
  \providecommand\BibTeX{{%
    \normalfont B\kern-0.5em{\scshape i\kern-0.25em b}\kern-0.8em\TeX}}}

\setcopyright{acmcopyright}
\copyrightyear{2020}
\acmYear{2020}

\acmConference{KDD-MLF-2020}{August 24, 2020}{San Diego, CA, USA}
\begin{document}

\title{Alpha Discovery Neural Network, the Special Fountain of Financial Trading Signals}
\author{Jie Fang}
\authornote{First Author}
\affiliation{%
  \institution{Tsinghua Shenzhen International Graduate School}
  \city{Shenzhen}
  \country{China}}
\email{fangx18@mails.tsinghua.edu.cn}

\author{Shutao Xia}
\authornote{Corresponding Author}
\affiliation{%
  \institution{Tsinghua Shenzhen International Graduate School}
  \city{Shenzhen}
  \country{China}}
\email{xiast@sz.tsinghua.edu.cn}

\author{Jianwu Lin}
\affiliation{%
  \institution{Tsinghua Shenzhen International Graduate School}
  \city{Shenzhen}
  \country{China}}
\email{lin.jianwu@sz.tsinghua.edu.cn}

\author{Zhikang Xia}
\affiliation{%
  \institution{Tsinghua Shenzhen International Graduate School}
  \city{Shenzhen}
  \country{China}}
\email{xzk19@mails.tsinghua.edu.cn}

\author{Xiang Liu}
\affiliation{%
  \institution{Tsinghua Shenzhen International Graduate School}
  \city{Shenzhen}
  \country{China}}
\email{liuxiang19@mails.tsinghua.edu.cn}

\author{Yong Jiang}
\affiliation{%
  \institution{Tsinghua Shenzhen International Graduate School}
  \city{Shenzhen}
  \country{China}}
\email{jiangy@sz.tsinghua.edu.cn}

\begin{abstract}
Genetic programming (GP) is the state-of-the-art in financial automated feature construction task. It employs reverse polish expression to represent features and then conducts the evolution process. However, with the development of deep learning, more powerful feature extraction tools are available. This paper proposes Alpha Discovery Neural Network (ADNN), a tailored neural network structure which can automatically construct diversified financial technical indicators based on prior knowledge. We mainly made three contributions. First, we use domain knowledge in quantitative trading to design the sampling rules and object function. Second, pre-training and model pruning has been used to replace genetic programming, because it can conduct more efficient evolution process. Third, the feature extractors in ADNN can be replaced by different feature extractors and produce different functions. The experiment results show that ADNN can construct more informative and diversified features than GP, which can effectively enriches the current factor pool. The fully-connected network and recurrent network are better at extracting information from the financial time series than the convolution neural network. In real practice, features constructed by ADNN can always improve multi-factor strategies$'$ revenue, sharpe ratio, and max draw-down, compared with the investment strategies without these factors.
\end{abstract}



\maketitle

\section{Introduction}
In quantitative trading, predicting the future return of stocks is one of the most important and challenging tasks. Various factors can be used to predict the future return of stocks. Such as the history price, volume and company$'$s financial data, etc. Normally, researchers define the features constructed from price and volume as technical indicators, the features constructed from the company$'$s financial data as fundamental data. On this task, various famous multi-factor models have been proposed, and many classical technical and fundamental factors have been constructed. For example, Fama-French Three-Factor Model ~\cite{FAMA19933} leverages three important factors that can provide the majority of information to explain the stock return. Later on, there are Fama-French Five Factor Model ~\cite{RePEc:eee:jfinec:v:116:y:2015:i:1:p:1-22}, and many other factors constructed by human experts. However, there are two shortcomings. First, it$'$s very expensive to hire human experts. Second, a human can$'$t construct some nonlinear features from high dimension data. Thus, both academic researchers and institutional investors have paid more and more attention to automated financial feature construction task~\cite{motoda2002feature}.

Feature construction is a process that discovers the relationships between features, and augments the space of features by inferring or creating new features. In this process, new features can be generated from a combination of existing features~\cite{liu1998feature}. A more straightforward description is that the algorithms use operators, hyper-parameters and existing features to construct a new feature. Sometimes both feature construction and feature selection can be merged together in one procedure. These methods consist of the wrapper, filtering, and embedded~\cite{chandrashekar2014survey}. Filtering is easy but achieves poor performance; it utilizes only some criteria to choose a feature and sometimes it can help us to monitor the feature construction process. Wrapper performs well by directly applying the model$'$s results as an object function. Thus, it can treat an individually trained model as a newly constructed feature. However, a considerable amount of computational resources and time are required. Embedded is a method that uses generalized factors and a pruning technique to select or combine features, which serves as a middle choice between filtering and wrapper. The most well-known and frequently employed automated feature construction method is Genetic Programming (GP), which is a kind of wrapper method that reverses polish expression to represent features and then simulates the evolution process. However, different domains require different object functions, and the input data$'$s data structure may differ~\cite{krawiec2002genetic}. Thus, it$'$s very important to do this task within a specific domain. This method has been shown to work well in many industries, such as object detection~\cite{kwakkenbos2010generation}, finance~\cite{romero2004knowledge}, and database management~\cite{watts2007mining}. However, the drawback of the method is that the constructed formulas are very similar and may cause co-linearity. In the financial feature construction task, the benchmark is genetic programming algorithm. It uses genetic programming to conduct the evolution process of formulaic factors ~\cite{allen1999using}~\cite{thomas1999importance}. WorldQuant ~\cite{RePEc:arx:papers:1601.00991} made public 101 formulaic alpha factors, which are also constructed by using this method. However, this method didn$'$t produce diversified features. The constructed features are similar, and they didn$'$t contain a very high level of information.

With the development of deep learning, more and more researchers begin to use the neural network to extract features from raw data and then add a fully-connected layer to reshape the feature$'$s output. Similarly, a trained model represents a newly constructed feature. Yang Zhong~\cite{zhong2016face} leverages it on pattern recognition tasks, he employs a CNN model to construct facial descriptors, and this method produces features that have considerably more information than the past method. K Shan~\cite{shan2017automatic} conducts experiments on this task and employs a deeper and wider convolution neural network. Hidasi B~\cite{hidasi2016parallel} uses a recurrent neural network to pre-locate the feature-rich region and successfully constructs more purified features. In a text classification task, Botsis T~\cite{botsis2011text} leverages recurrent neural networks to build a rule-based classifier among text data, in which each classifier represents a part of the text. S Lai~\cite{lai2015recurrent} proposes a network structure that uses both a recurrent neural network and a convolution neural network to extract text information. With the help of a neural network$'$s strong fitting ability, we can produce highly informative features by tailoring the network structure for different industries. In financial feature construction tasks, researchers have begun to use a neural network to give an embedding representation of financial time series. More specifically, Fuli Feng~\cite{RePEc:arx:papers:1810.09936} leverages LSTM to embedding various stock time series, and then uses adversarial training to make a binary classification on stock$'$s future return. Leonardo~\cite{RePEc:arx:papers:1912.07700} adopts well designed LSTM to extract features from unstructured news data, and then form a continuous embedding. The experiment result shows that these unstructured data can provide much information and they are very helpful for event-driven trading. Zhige Li ~\cite{RePEc:wsi:ijitmx:v:04:y:2007:i:02:n:s0219877007001016} leverages a Skip-gram architecture to learn stock embedding inspired by a valuable knowledge repository formed by fund manager$'$s collective investment behaviors. This embedding can better represent the different affinities over technical indicators. With a similar idea, we use a neural network to give a brief embedding of long financial time series. This embedding can help to summarize the most important information in the high dimension data. Different from the previous work, we mainly make three contributions in this paper. First, we strictly design the sampling rules. All the stocks on the same trading day are regarded as one batch, which meets economic principles. Second, we didn$'$t simply use the stock return to serve as object function, but we use the spearman coefficient of stock$'$s return and stock$'$s feature value to serve as object function. We are the first to use this object function in neural network and we also have fixed its un-derivable problem. Third, we adopt pre-training and model pruning to add up enough diversity into our constructed features, which helps this system to produce more diversified features than the benchmark.
In this paper, we proposed a novel network structure called ADNN, which is tailored for stock time series. This framework can use different deep neural networks to automatically construct financial factors. ADNN has outperformed the benchmark on this task, from the perspective of all frequently compared indicators. What$'$s more, we find some interesting differences between different feature extractors on this task, and we conduct experiments to comprehend them.

\section{ALGORITHM INTRODUCTION}
\subsection{Benchmark}
In quantitative trading, investors commonly construct factors, and regard these factors as trading signals. In automated financial feature construction task, what we want is to let a algorithm to automatically construct new factors, to determine the variable, operator and hyper-parameters.

The benchmark on this task is GP. It uses a reverse polish expression to represent the feature$'$s formula and then leverages binary tree to store its explicit expression. In each training iteration, researchers leverage GP to conduct the evolution process. This evolution process includes merging different formulas, cutting some parts of the formula and changing some parts of the formula, etc. The training process is shown in Figure 1.

\begin{figure}[htbp]
\centering
\vspace{-0.3cm}
\includegraphics[width=8.5cm]{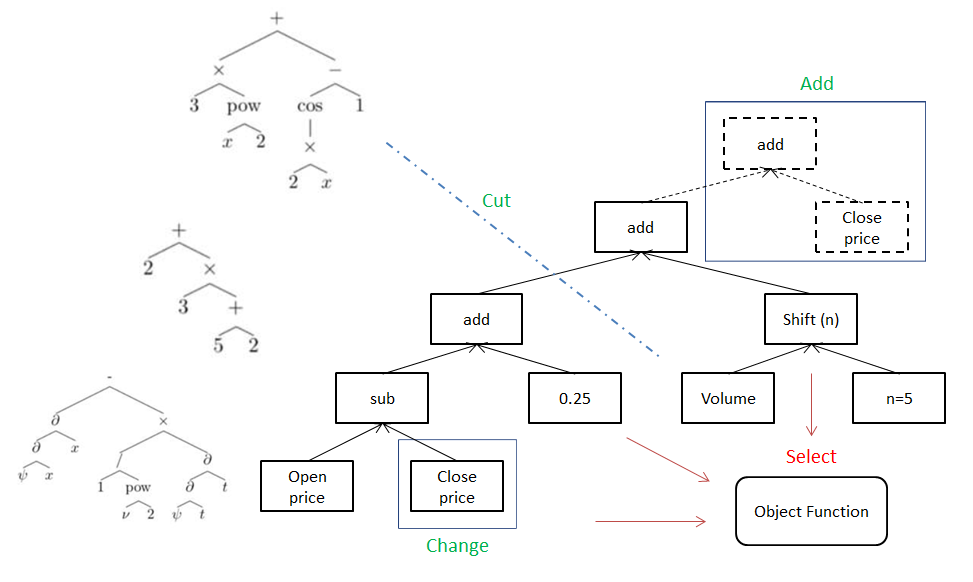}

\caption{This is the GP$'$s evolution process. Each tree represent a formulaic factors, and the right tree will get survived according to the object function.}
\vspace{-0.2cm}
\end{figure}

As shown in Figure 1, researchers add diversity into the constructed features by changing a part of the reverse polish expression. For example, we have a frequently used factor 1, shown in formula 1. And then we make a small change on factor 1, in order to construct a new factor, shown in formula 2.

\begin{equation}
Factor1=\frac{high\,price-low\,price}{volume.shift(1)}
\end{equation}

\begin{equation}
Factor2=\frac{high\,price-volume}{volume.shift(1)}
\end{equation}

Factor 1 means the relative strength of price compared with volume, which has economic meaning. However, factor 2 is totally different from factor1, and it is really hard to explain. Because in this algorithm, the parent factor and child factor have little in common. The parent factor has high IC, but the child factor may not successfully inherent the good characteristics from its parent factor. As a result, we think GP is not a good method to construct new factors, due to its low efficient evolution process on this task.

\subsection{Alpha Discovery Neural Network}
The network structure of the ADNN is shown in Figure 2. The major contributions of this novel network structure includes 1). ADNN uses Spearman Correlation to serve as a loss function, which mimics human practices of quantitative investment. And the sampling rules also meet economic principle. 2). A meaningful derivable kennel function is proposed to replace the un-derivable operator $rank()$. 3). We use pre-training and pruning to replace the GP$'$s evolution process, which is more efficient.

\begin{figure}[htbp]

\centering

\includegraphics[width=8.5cm]{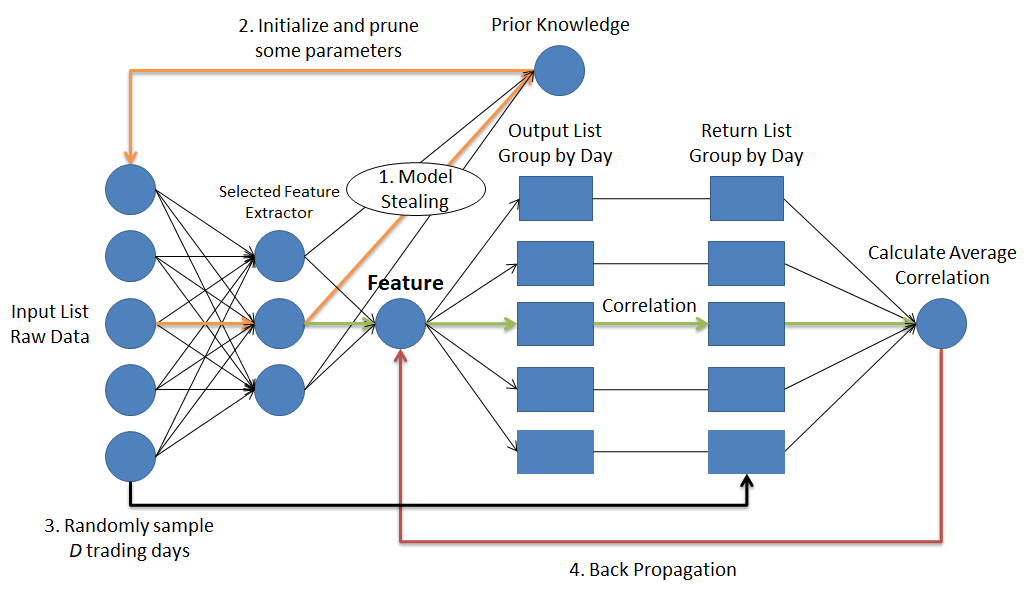}

\caption{Alpha discovery neural network$'$s structure.}

\end{figure}

As shown in Figure 2, in each back-propagation, ADNN randomly samples $D$ trading days$'$ data, and then calculate the Spearman Coefficient of factor value and factor return in each trading day. $D$ should be larger than 3, and taking $D$ trading day$'$s information into account can help the neural network to get a more stable convergence. Quantitative investors care more about the relative strength of each stock on the same trading day, rather than its absolute strength. Thus, doing calculation in each trading day and using the Spearman Coefficient to serve as loss function is reasonable.

In each batch, we assume that there are $m$ stocks that belong to this trading day. The input tensor$'$s shape is $(m,5,n)$, because there are $m$ samples, and 5 types of time series, which is the open price, high price, low price, close price, and volume. Each time series$'$ input length is $n$. We also name the output tensor as factor value, with shape $(m,1)$. The factor return tensor$'$s shape is $(m,1)$, which means the revenue that we can earn from this asset for a long period of time. The length of the holding time is $a$. Here, we assume that all the feature extractors in Figure 2 are Multi-layer Perceptron (MLP), which is easy for us to give a general mathematics description. In the experiment part, we will show the experiment results based on more complicated and diversified feature extractors. $w_{i}$ means the kernel matrix in $i\,th$ layer, $b_{i}$ means the bias matrix in $i\,th$ layer, $a_{i}$ means the activate function in $i\,th$ layer, and there will be $p$ layers in total.

\begin{equation}
  \begin{split}
  x&=l_{p}=a_{p}({w_{p}}^{T}l_{p-1}+b_{p}),\\
  l_{1}&=a_{1}({w_{1}}^{T}Input+b_{1}).
  \end{split}
  \end{equation}

\begin{equation}
y=Factor\,Return=\frac{close \, price_{t+a}}{close \, price_{t}}-1
\end{equation}

We apply a Spearman Correlation to calculate the correlation between a factor value and a factor return. This setting can help us to obtain powerful features that are suitable to forecast the future stock return. And this setting also makes our batch size and sampling rules become meaningful. Only the data belongs to the same trading day, should be involved in the same batch. However, Spearman Correlation uses operator rank() to gid rid of some anomalies in financial time series. Rank() is not derivable, which is not acceptable for the neural network. Thus, we use a derivable kernel function g(x) to replace rank().

\begin{equation}
g(x)=\frac{1}{1+exp(-p*\frac{x-\bar{x}}{2*std(x)})}
\end{equation}

As shown in formula 5, at first, it projects x into a normal distribution which is zero-centralized. Next, it uses a hyper-parameter p to make sure that the 2.5\%-97.5\% of data should lay in the range between $[mean-2std,mean+2std]$. Thus, p equals to 1.83. we can get p=1.83. For example, one out-lier $x_i=\bar{x}+2std(x)$, and $\frac{g(x_i)-g(\bar{x})}{g(\bar{x})}\leq\frac{x_i-\bar{x}}{\bar{x}}$, so the result is $std\leq0.362\bar{x}$. It means if one distribution$'$s standard deviation is large, and it is larger than $0.362\bar{x}$, the g(x) can shorten the distance between outliers and the central point. If the distribution$'$s standard deviation is very small, g(x) will make it worse. However, even in this case, we can make sure that 95\% of the points are between $[mean-2std,mean+2std]$, which is acceptable. The final object function is defined in formula 6, where E(x) represents the expected value of x, $\bar{x}$ represents the average value of x. And in each batch, we calculate the average value from $q$ trading days, which can make the optimization process more stable.

\begin{equation}
\begin{split}
IC(x,y)&=\frac{E(g(x)-\bar{g(x)},g(y)-\bar{g(y)})}{E(g(x)-\bar{g(x))}E(g(y)-\bar{g(y)}))},\\
Loss&=-\frac{1}{q}\sum_{i=1}^{q}IC(x_{i},y_{i}).
\end{split}
\end{equation}

\subsection{Put prior knowledge into network}
Combining with model stealing~\cite{juuti2019prada} and pruning on input data can improve the signal$'$s diversity. Model stealing means that if the input $x$ and the output $y$ are known, our network can obtain a suitable parameter $w$ to fit this projection. However, this technique is not always helpful to learn a distribution without tailoring the network structure. If we have a fixed network structure, and we have no idea about the target distribution, the techniques such as removing the outliers, will be very helpful for the continuous prior knowledge. Using high-temperature T also works for the discrete prior knowledge.

Pre-training uses $f(x)=a(w^T x+b)$ to embed the input data (the data is embedded by MLP (Several fully-connected layers with tanh and relu activation functions. The number of neural in each layer should be decided by the length of input time series), $w$ means kernel matrix, $b$ means bias matrix, $a$ means activation function) and then use this embedded layer to mimic the prior knowledge. In this part, we use the mean squared error as the object function.

\begin{equation}
    \mathop{\arg\min}_{a,b,w} \ \ \frac{1}{n}\sum_{i=1}^N (y_i-f(x_i))^{2}
\end{equation}

Almost all technical indicators can be easily learned by using MLP. Here, MSE or MAE can$'$t represent the real pre-training performance, because all factor values are really small, which makes all MSE value very small. In order to have a better measurement of the performance, $\frac{1}{n}\sum_{i=1}^N |\frac{y_i-f(x_i)}{y_i}|$ is used to measure its error rate. Some classical technical indicators, such as MA, EMA, MACD, RSI, BOLL, and other typical financial descriptors are selected as prior knowledge for pre-training, shown in Table 1.

\begin{table}[htbp]
\centering
\caption{Here are the formula of some classical technical indicators and financial descriptors. They serve as prior knowledge for ADNN. Close refers to stock close price, volume refers to stock volume, and AdjClose refers to adjusted close price.}
\begin{tabular}{cl}
\hline
Technical Indicator  & Mathematical Expression\\
\hline
\multirow{1}{*}{MA}
&$MA_{N}(x_{n})=\frac{1}{N}\sum_{k=0}^N x_{n-k}$\\
\hline
\multirow{1}{*}{EMA}
&$EMA_{N}(x_{n}) = \frac{2}{N+1} \sum_{k=0}^\infty(\frac{N-1}{N+1})^{k}x_{n-k}$\\
\hline
\multirow{1}{*}{MACD}
&$MACD = EMA_{m}(i) -EMA_{n}(i)$\\
\hline
\multirow{2}{*}{PVT}
&$PVT(i) = PVT(i-1)+volume(i)*$\\
&$(close(i)-close(i-1))/close(i-1)$\\
\hline
\multirow{2}{*}{TOP10}
&$MA10 = MA\left(Close,10\right)$\\
&$TOP10 = \frac{MA10}{MA10_{top10\%}}-1$\\
\hline
\multirow{4}{*}{DC}
&$H = MA\left( High \times AdjClose/Close,n\right)$\\
&$L = MA\left( Low \times AdjClose/Close,n\right)$\\
&$M = \frac{1}{2}\left(H + L\right)$\\
&$DC = AdjClose / M$\\
\hline
\multirow{5}{*}{BOLL}
&$StdV = MStdv\left(Close,n\right)$\\
&$Mean = MA\left(Close,n\right)$\\
&$LB = Mean - Stdv$\\
&$BBL = \frac{LB}{Close}$\\
&$MStdv_{n,t} = Stdv\left(Close_{t-n:t}\right)$\\
\hline
\end{tabular}
\label{tab:plain}
\end{table}

Some descriptors with different parameters such as DC(5) and DC(15) will be regarded as different prior knowledge because they have given enough diversity to ADNN. The testing error rate of pre-training these factors, shown in Table 1, is $0.081 \pm 0.035$. We think this error rate is acceptable, and it can bring enough diversity into the network.

Why is pre-train with prior knowledge needed? Because knowledge is the source of diversity, we should keep it. According to the concept of Muti-task Learning, pre-training can keep some part of the domain knowledge in the network. In order to keep more diversity after the pre-training process, pruning is needed. Permanently pruning the useless elements in the embedding matrix can help us to keep the diversity, and filter out noisy signals from prior knowledge. High pruning rate will lose too much information, but low-level pruning rate is hard to keep the diversity. The ideal pruning rate should be about 0.2-0.5, which means 20\%-50\% of the elements in the mask matrix should be 0. All the setting is the same as ~\cite{frankle2018lottery}, and here are more explanations. After embedding the data as f(x), we get its parameter matrix $w$. Then we create a mask matrix $m$ to prune the parameters. For example, $w_{ij}$ in the parameter matrix is relatively small, which means this element is useless, and we should set $m_{ij}$=0 to prune it. If the $w_{ij}$ is not useless, then we set $m_{ij}$=1. This method can help us to further keep the diversity in the neural network, and let the network focus on improving the current situation. The pruning process is shown in formula 8, here $m*w$ means the Hadamard Product.

\begin{equation}
    f(x)=(m*w)^\mathrm{ T }x+b
\end{equation}

After pre-training and pruning the network, we use the object function shown in formula 6 to train ADNN. We simply reshape the input data into a picture. And then we use the Saliency Map to look at how the raw data contribute to the final constructed factor. The training process is shown in Figure 3, the y-axis is [open price, high price, low price, close price, volume], the x-axis is the length of input time series.

\begin{figure}[htbp]

\centering

\includegraphics[height=5.5cm,width=8.4cm]{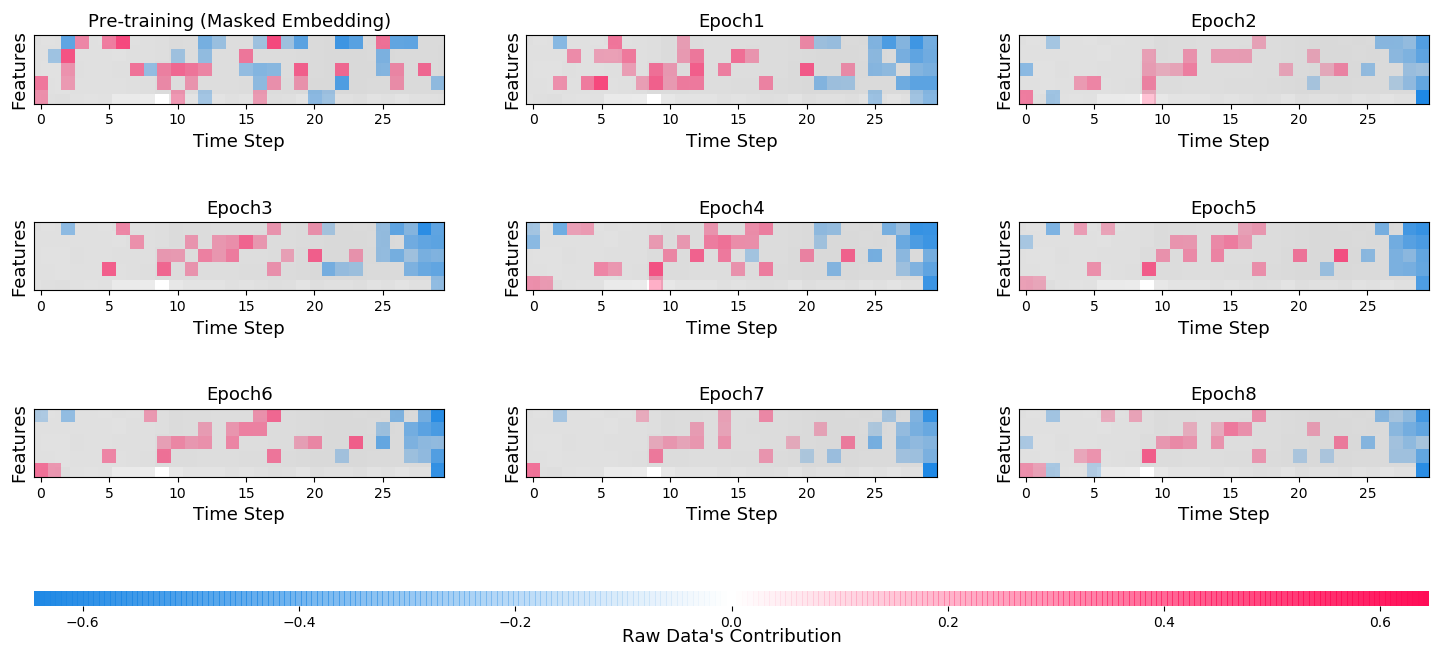}

\caption{How ADNN leads the prior knowledge to become a better technical indicator. ADNN can adjust the raw data$'$s contribution according to the objective function, and make it better compared with its initial state.}

\end{figure}

Prior knowledge performs like a seed, which is the source of diversity in this system. Although the features constructed by ADNN is not explicit, and it$'$s hard to explain, compared with GP. However, there are some strong points of ADNN. As mentioned above, ADNN can conduct a more efficient evolution process. After warming up the system, we can know how many differences have been put into the constructed factors. Second, although we can$'$t fully understand its formular, we at least know that whether this factor is momentum or reverse. For example, $close[t]-close[t-1]$ is a momentum factor, but $close[t-1]-close[t]$ is a reverse factor. Third, unlike the traditional factors, whose raw data$'$s contribution is discrete. The raw data$'$s contribution in ADNN is continuous, which helps it to extract high dimension information. Human experts can$'$t construct factors by extracting high dimension data, this huge differences can help to avoid factors crowding. After all, the trading opportunities is limited, people can$'$t share the same trading signals.~\cite{GREEN2019236}.

We conduct experiments on different feature extractors. There are two motivations to conduct experiments on different feature extractors. First, different feature extractors require different input data$'$s data structure. After performing a literature review and consulting professional experts in the market, we discover many different ways to organize the input data. However, none of them can prove that their structure is the best. Thus, experiments on these structures should be performed. The second motivation is that different extractors have their own strong comings and short comings. Some of them aim at extracting temporal information but the others aim at spatial information. Some of them designed for a long time series, but some of them are designed for quick training. We think all these differences can make our factor pool more diversified.

\section{Experiments }
\subsection{Experiment setting}
We use daily trading data in the Chinese A-share stock market (in the following part, we call it A-share market data), including the daily open price, high price, low price, close price and trading volume, in the past 30 trading days. The raw data is standardized by using its time-series mean and standard deviation in the training set. Both the mean and standard deviation are calculated from the training set. We attempt to use these inputs to predict the stock return in the next 5 trading days (using 3-15 trading days is recommended). Moreover, we should obey the market policy when we form a trading strategy.

We have done a lot of experiments to select reasonable hyper-parameters. For each experiment, 250 trading days serve as the training set, the following 30 trading days serve as the validation set, and the following 90 trading days serve as the testing set. The constructed factors can keep high IC during the next 90 trading days. Most importantly, we want to stress a counter-intuitive setting. The training period should be no longer than 250 trading days because financial features are non-stationary. If we request a feature that can work well for a very long period of time, then we will only find this feature in the over-fitting situation. Thus, we design a rolling forecast structure that we will automatically find powerful features for each trading day. Each automatically constructed features will have their own highlight time on this trading day. What$'$s more, these factors not only work well on this single day. Actually, his performance can last several trading days, with a moderate decay.

To make a fair comparison, the same setting is deployed for the GP algorithm. This algorithm$'$s logic references relative work~\cite{thomas1999importance} and~\cite{allen1999using}. Besides, the input data$'$s period and type should be the same. In this paper, we analyze the construed features$'$ performance from different perspectives. Normally, institutional investors use Information Coefficient (IC), shown in formula 6, to measure how much information carried by a feature. For diversity, the cross-entropy is used to measure the distance between two different features$'$ distributions on the same trading day.

\begin{equation}
Distance(f_1,f_2)=\sum softmax(f_1) log \frac{1}{softmax(f_2)}
\end{equation}

In formula 9, $f_1$ and $f_2$ refers to different features$'$ distribution in the same trading day. The softmax function can help us get rid of the effect from scale without losing its rank information. And k-means is used to cluster the distance matrix of this relative distance between two features. The average distance between each cluster center refers to the diversity of this algorithm on this trading day. Besides measurements of IC and diversity, the performance of a trading strategy based on the constructed features are also measured, such as absolute return, max-drawdown~\cite{cvitanic1994portfolio}, and sharp-ratio~\cite{sharpe1994sharpe}. Basically, all these indicators are really important to assess a feature$'$s performance.

\subsection{Beat the state-of-the-art technique}
The network structure shown in Figure 2 can equip ADNN with different deep neural networks. In order to show the general situation, we equip ADNN with 4 fully-connected layers. Each layer has 128 neural, tanh activate function, L2 Regularization, and dropout technic. This general and simple setting is enough to beat the GP. We put forward three schemes help to show how ADNN beat the GP. \emph{Only GP} means only using genetic programming, \emph{Only ADNN} means only use ADNN to construct factors, \emph{GP\&ADNN} means use GP$'$s value to initialize ADNN and then construct factors. All the experiments are conducted out of the sample, and we have summarized it in Table 2.

\begin{table}[htbp]
\caption{The performance of different schemes.}
\centering
\begin{tabular}{ccc}
\hline
Object  & Information Coefficient & Diversity \\
\hline
\emph{Only GP}       & 0.094  & 17.21     \\
\emph{GP\&ADNN}       &0.122  & 25.44      \\
\emph{Only ADNN}    & 0.107  & 21.65     \\
\hline
\end{tabular}
\label{tab:plain}
\end{table}

As shown in table 2, \emph{Only ADNN} is better than \emph{Only GP}, which means ADNN outperforms GP on this task. And we also find that \emph{GP\&ADNN} is the best, it means that our method can even improve the performance of GP.

In real practice, we should leverage the constructed factors to form a multi-factor strategy and compare its performance with GP. The specific strategy setting is same as section 3.4, and we have repeated this experiment on different periods of time. The long-term backtest result is shown in Table 3, \emph{Only ADNN} always has better performance than the \emph{Only GP}. It shows that ADNN has also beaten the SOTA in real practice. Similar to the conculsions made above, if we combine these two methods together, the combined factors$'$ strategy has the best performance in backtesting.

\begin{table}[htbp]
\centering
\caption{Strategy$'$s absolute return for each scheme.}
\resizebox*{8.0cm}{3.3cm}{
\begin{tabular}{ccccc}
\hline
Time                                                                                  & \emph{Only GP} & \emph{GP}\&\emph{ADNN} & \emph{Only ADNN} & \emph{ZZ500}    \\
\hline
\begin{tabular}[c]{@{}c@{}}Train:2015.01-2015.12\\ Test: 2016.02-2016.03\end{tabular} & +2.59\%  & +5.74\%  & +4.52\%  & +1.67\%  \\
\hline
\begin{tabular}[c]{@{}c@{}}Train:2016.01-2016.12\\ Test: 2017.02-2017.03\end{tabular} & +5.40\%  & +10.26\% & +8.33\%  & +2.53\%  \\
\hline
\begin{tabular}[c]{@{}c@{}}Train:2017.01-2017.12\\ Test: 2018.02-2018.03\end{tabular} & -5.27\%  & -4.95\%  & -4.16\%  & -6.98\%  \\
\hline
\begin{tabular}[c]{@{}c@{}}Train:2018.01-2018.12\\ Test: 2019.02-2019.03\end{tabular} & +13.00\% & +15.62\% & +15.41\% & +13.75\% \\
\hline
\end{tabular}
}
\label{tab:my-table}
\end{table}

All the results shown above is based on the most basic feature extractors. So will there be more powerful feature extractors to discover knowledge from financial time series? And what is the suitable input data structure for financial time series?

\subsection{Comparing different feature extractors}
All experiments are conducted in the same setting mentioned in section 3.1, and the results are summarized after generating 50 features. For the hardware equipment, we use 20 g GPU (NVIDIA 1080Ti) and 786 g CPU (Intel Xeon E5-2680 v2, 10 cores). Based on this setting, we show the amount of time that we need to train 50 neural networks. Moreover, the time to restore 50 trained networks and obtain their feature values will be substantially faster than traditional features. Because most traditional features are constructed with complicated explicit formulas, these formulas are not suitable for matrix computing. Using neural networks to represent features in matrix computing, which can have a faster testing speed.

\begin{table}[htbp]
\centering
\caption{The higher are the information coefficient (IC) and diversity, the better is their performance. Normally, a good feature$'$s long-term IC should be higher than 0.05, but it cannot be higher than 0.2 in an A-share market.}
\setlength{\tabcolsep}{1.0mm}{
\begin{tabular}{ccccc}
\hline
Type  & Network & IC & Diversity & Time\\
\hline
Baseline &GP& 0.072  &17.532 &0.215 hours    \\
\hline
Vanilla &FCN &0.124 &22.151  &0.785 hours           \\
\hline
\multirow{2}{*}{Spatial}
& \multicolumn{1}{c}{Le-net} &0.123 &20.194 &1.365 hours \\
&Resnet-50 &0.108 &21.403 &3.450 hours \\
\hline
\multirow{3}{*}{Temporal}
&LSTM &0.170 &24.469 &1.300 hours \\
&TCN &0.105 &21.139  &2.725 hours           \\
&Transformer &0.111 &25.257 &4.151 hours    \\
\hline
\end{tabular}}
\label{tab:plain}
\end{table}

Shown in Table 4, basically, all neural networks can produce more diversified features than using GP. But temporal extractors are especially better at producing diversified features, such as LSTM~\cite{hochreiter1997long} and Transformer~\cite{vaswani2017attention}. As for TCN~\cite{lea2017temporal}, the author who put forward this network structure proves its ability to capture the temporal rules buried in data. However, there is a huge difference. TCN relies on a convolution neural network, but LSTM and Transformer still contain recurrent neural networks (Normally, the transformer uses a recurrent neural network to embedded the input data). The existence of a recurrent neural network structure may contribute to the difference in diversity. For Le-net~\cite{lecun1998gradient} and Resnet~\cite{he2016deep}, they don$'$t provide us with more informative features. It looks like that the convolution network structure is not suitable to extract information from the financial time series.

\subsection{Real-world use case}
In real practice, we combines traditional factors and the factors constructed by ADNN to form a quantitative investment strategy. What we want is to see if ADNN can enrich the factor pool and improve the traditional multi-factor strategy.

We form a frequently used multi-factors strategy to test its performance in the real case. In the training set, the sample whose return ranked in the top 30\% in each trading day is labeled as 1 and the sample whose return ranked in the last 30\% of each trading day is labeled as 0. We abandon the remaining samples in the training set~\cite{FAMA19933}. After training these features with XGBoost~\cite{chen2015xgboost} using binary logistics mode, the prediction result reflects the odds that this stock has outstanding performance in the following 5 trading days. It defines the 50 features constructed by human experts as \emph{PK 50}, the features constructed by ADNN as \emph{New 50}, and the features constructed by both GP and PK as \emph{GP-PK 50}. In separate experiments, we use XGBoost to pre-train both \emph{PK 50} and \emph{New 50} in the training set and then using the weight score from XGBoost to choose the 50 most important features as \emph{Combined 50}. This feature selection process only happens once, and only be conducted in training set.

\begin{table}[htbp]
\centering
\caption{Back testing starts from Jan 2019 to June 2019. The investment target is all A-share, except for the stock can$'$t be traded during this period of time. Strategy$'$s commission fee is 0.5\%. SR refers to Sharpe Ratio, MD represents Max-Drawdown.}
\resizebox*{8.5cm}{7.5cm}{
\begin{tabular}{cccccc}
\hline
Type                                          & Target                       & Group       &Revenue  &MD & SR\\ \hline
\multicolumn{1}{l}{\multirow{5}{*}{Baseline}} & ZZ500                        & Stock Index & 19.60\% & 13,50\%       & 1.982        \\ \cline{2-6}
\multicolumn{1}{l}{}                          & HS300                        & Stock Index & 18.60\% & 20.30\%      & 1.606        \\ \cline{2-6}
\multicolumn{1}{l}{}                          & PK                           & PK 50       & 24.70\% & 18.90\%      & 2.314        \\ \cline{2-6}
\multicolumn{1}{l}{}                          & \multirow{2}{*}{GP}          & GP 50       & 17.60\% & 25.30\%      & 1.435        \\ \cline{3-6}
\multicolumn{1}{l}{}                          &                              & GP-PK 50    & 25.40\% & 14.80\%      & 2.672        \\ \hline
\multirow{2}{*}{Vanilla}                      & \multirow{2}{*}{FCN}         & New 50      & 20.60\% & 15.80\%      & 2.189        \\ \cline{3-6}
                                              &                              & Combined 50 & 29.60\% & 15.70\%      & 3.167        \\ \hline
\multirow{4}{*}{Spatial}                      & \multirow{2}{*}{Le-net}      & New 50      & 18.00\% & 16.90\%      & 1.800        \\ \cline{3-6}
                                              &                              & Combined 50 & 27.50\% & 16.40\%      & 2.921        \\ \cline{2-6}
                                              & \multirow{2}{*}{Resnet-50}   & New 50      & 19.90\% & 15.40\%      & 1.962        \\ \cline{3-6}
                                              &                              & Combined 50 & 29.30\% & 17.20\%      & 2.787        \\ \hline
\multirow{6}{*}{Temporal}                     & \multirow{2}{*}{LSTM}        & New 50      & 19.50\% & 13.00\%      & 2.205        \\ \cline{3-6}
                                              &                              & Combined 50 & 29.90\% & 15.00\%      & 3.289        \\ \cline{2-6}
                                              & \multirow{2}{*}{TCN}         & New 50      & 22.40\% & 14.70\%      & 2.440        \\ \cline{3-6}
                                              &                              & Combined 50 & 26.90\% & 16.80\%      & 2.729        \\ \cline{2-6}
                                              & \multirow{2}{*}{Transformer} & New 50      & 21.10\% & 15.90\%      & 2.203        \\ \cline{3-6}
                                              &                              & Combined 50 & 27.20\% & 15.10\%      & 2.806        \\ \hline
\end{tabular}%
}
\end{table}

As shown in Table 5, \emph{HS300} and \emph{ZZ500} are important stock indices in the A-share stock market. Revenue represents the annualized excess return, by longing portfolio and shorting the index. The max drawdown is the worst loss of the excess return from its peak. The Sharpe ratio is the annually adjusted excess return divided by a certain level of risk. These indicators can show the strategy$'$s performance from the perspective of both return and risk.

For the \emph{New 50}, although they have higher IC than the \emph{PK 50}, their overall performance is not always better than \emph{PK 50}. Because the overall performance of a multi-factor strategy is determined by both diversity and information volume (IC), we guess the diversity of \emph{PK 50} is remarkably higher than the diversity of \emph{New 50}. We also did experiment to verify this guess. Thus, although every single new factor is better than the old factor, their overall performance not always be better. ADNN$'$s diversity is larger than the GP, but for further research, making ADNN$'$s diversity even larger is still badly needed. In the real world use case, all investors have their own reliable and secret factor pool, what they want is that the new constructed factors can bring in margin benefits. Thus, they will use both new and old factors to do trading. That$'$s the reason why \emph{Combined 50} can represent ADNN$'$s contribution in the real situation. In all cases, \emph{Combined 50} is better than \emph{PK 50} and \emph{GP-PK 50}, which means that the ADNN not only perform better than GP, but also can enrich investors$'$ factor pool. We also plots the exceed return curve of these strategies, shown in Figure 4.

\begin{figure}[htbp]
\vspace{-0.2cm}
\centering

\includegraphics[height=6.0cm,width=8.5cm]{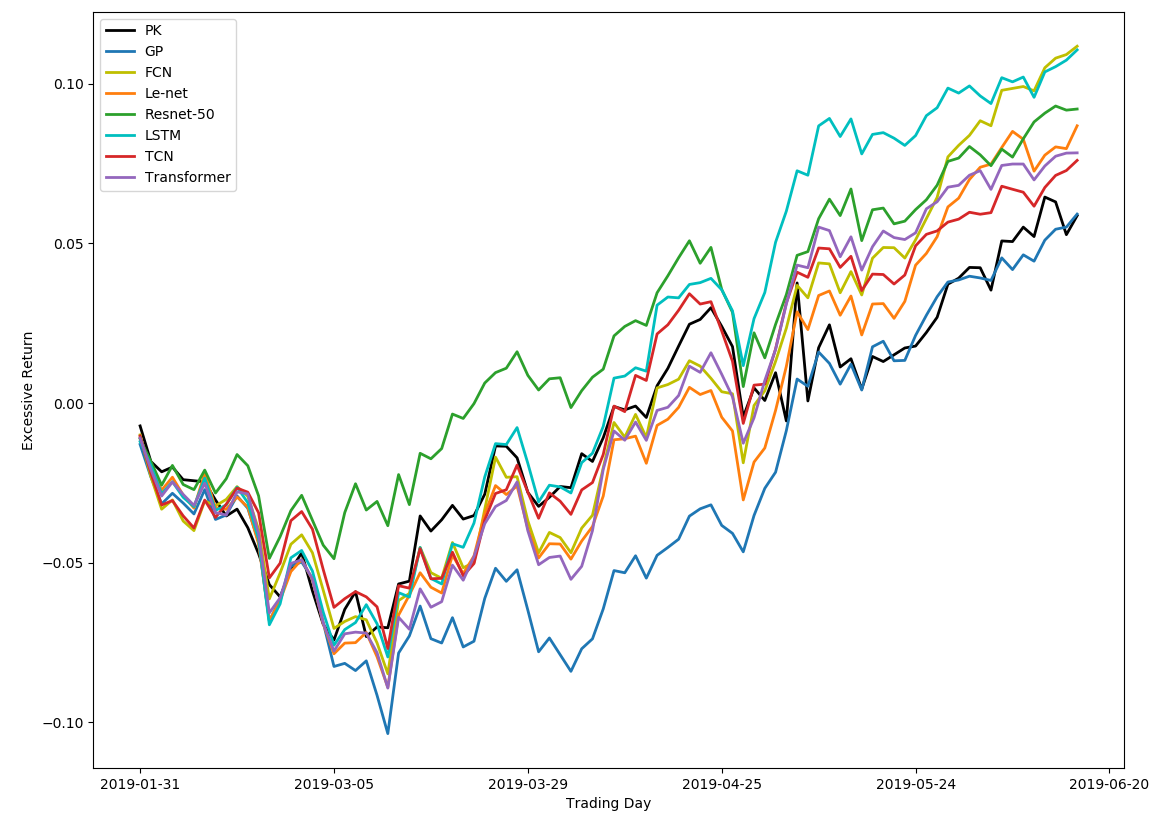}

\caption{Different feature extractors$'$ exceed return in testing set, hedge on HS300 Index.}
\vspace{-0.2cm}
\end{figure}

Shown in Figure 4, all these curves are similar, due to the fact that they all shared some factors from \emph{PK 50}. And all the schemes powered by ADNN performs better than GP. During this period of time, they have beaten the market more than 10 percent. It is reasonable because all the features are only constructed from price and volume data. They don$'$t contain any fundamental data or even sentiment data. What$'$s more, we will get a lot of extra information during the feature construction process. This information is helpful in the feature selection process. That$'$s the main reason why some wrapper methods will do feature selection and construction at the same time. For further research, the current structure can be improved to conduct both the feature construction and feature selection process at the same time. This paper directly leverages this reasonable and fair feature selection method, because it only focuses on the feature construction task.

\subsection{Comprehend the result}
From the experiment result, we have found that different feature extractors perform differently. In this part, we try to comprehend this result. We construct 50 features by using FCN, 50 features from the network focused on spatial information and 50 features from the network focused on temporal information. Then the diversity is clustered into three groups using k-means; this method has been mentioned in section 3.1. To show the distributions more clearly, we cluster them into three groups. Then we initialize one of the cluster centers at $(0,0)$ and then determine the other two cluster centers according to their relative distance and a given direction. This direction will only influence the outlook of this graph, but not influence the shared space between two different clusters. In the following experiments, we plot all the factors$'$ distributions to help us understand the characteristics of different types of feature extractors. Here, we focus on the sparsity and common area shared by each group. Because these two indicators can help us to comprehend which feature extractor really contributes, and how much special information it has discovered.

\begin{figure}[htbp]
\vspace{-0.2cm}
\centering

\includegraphics[height=4.15cm,width=8.3cm]{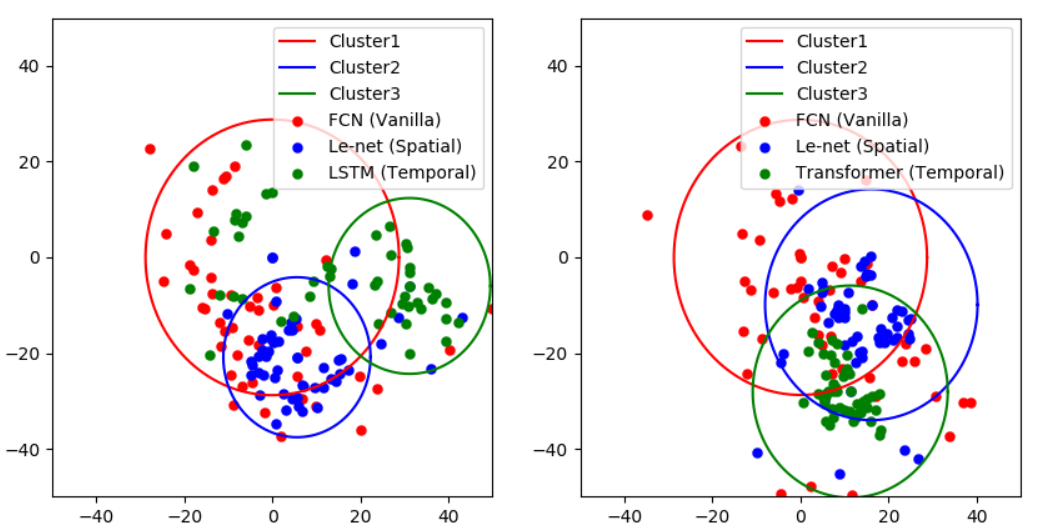}

\caption{Cluster different neural networks, spatial network against temporal network.}
\vspace{-0.1cm}
\end{figure}

As shown in Figure 5(left), the features constructed by the LSTM have the sparsest distribution, which means that the network structure that focuses on temporal information is excellent at extracting information from the financial time series. However, a large space is shared by FCN and Le-net. We can regard Le-net$'$s information as a subset of FCN. Combined with the convolution neural network$'$s poor performance in sections 3.2 and 3.3, it looks like that the convolution neural network structure does not have a substantial contribution to extracting information from the financial time series. Figure 5(right) is an extra experiment, whose result supports this conclusion as well.

\begin{figure}[htbp]
\vspace{-0.2cm}
\centering

\includegraphics[height=4.15cm,width=8.3cm]{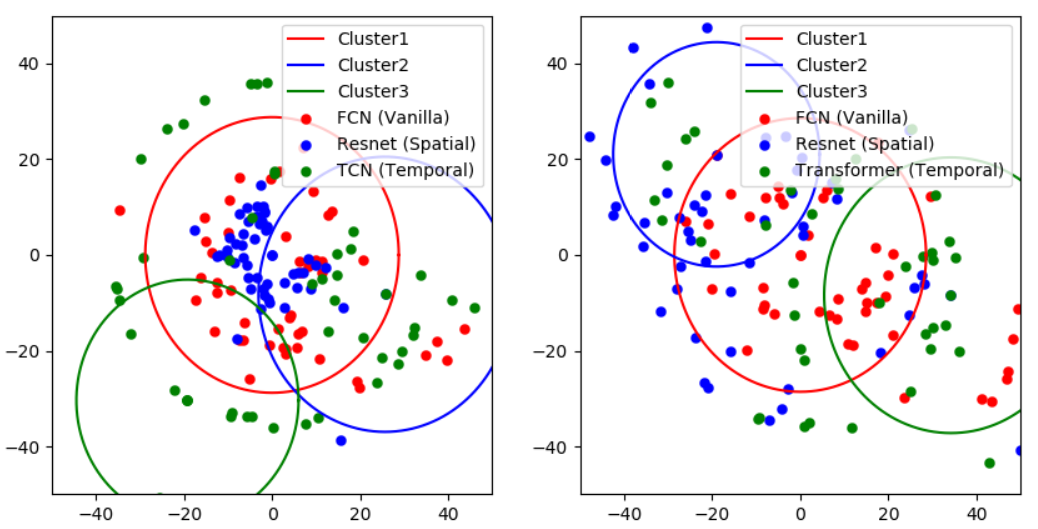}

\caption{Cluster different types of neural networks. More complicated, compared with the networks used in Figure 5.}
\vspace{-0.1cm}
\end{figure}

From Figure 6, we can draw the same conclusions as above. What$'$s more, the network used in Figure 5 is simple, but the network used in Figure 6 is relatively more complicated. Compare Figure 5 and Figure 6, we can find that the complex network takes larger space. It shows that complicated neural network bigger diversity. Thus, we think the complicated neural network$'$s strong point is that they have less possibility to get co-linearity. Commonly, the complex network has large parameter set. And at most of the time, its impressive performance comes from its large parameter set. A very complicated neural network will be helpful remember some stationary rules. But for the non-stationary stock market, the rules in training set maybe different from the rules in testing set. If we only rely on large parameter set to remember the rules, it may bring over-fitting risk. And currently, most of the tradings are still made by human, which means the majority of trading signals are still linear. Thus, at present, the very complicated neural network can$'$t have a promising performance in the stock market.

However, while the stock market is developing, more and more investors crowd into this game. We think that the factor crowding phenomenon will become more and more clear. In addition, as more and more trading is made by algorithms, the non-linear part in the trading signals will be larger. Thus, for quantitative trading, we believe that the complicated and tailored neural network structure will have its supreme moment in the near future.

\section{Conclusion}
In this paper, we put forward the alpha discovery neural network, which can automatically construct financial features from raw data. We designed its network structure according to the economic principle, and equip it with different advanced feature extractors. The numerical experiment shows that ADNN can produce more informative and diversified features than the benchmark on this task. In real practice, ADNN can also achieve better revenue, sharpe-ratio and max-drawdown than genetic programming. What$'$s more, different feature extractors play different roles. We have done plenty of experiments to verify it and try to comprehend its function. For further research, we will leverage this framework to automatically construct useful features based on the companies$'$ fundamental data and sentiment data.

\begin{acks}
This work is supported in part by the National Natural Science Foundation of China under Grant 61771273 and the R\&D Program of Shenzhen under Grant JCYJ20180508152204044.
\end{acks}

\bibliographystyle{ACM-Reference-Format}
\bibliography{reference}

\end{document}